\documentclass[a4paper,11pt,twoside]{article}

\usepackage{amsmath}
\usepackage{amssymb}
\usepackage{epsfig}
\usepackage{wrapfig}
\usepackage{array}
\usepackage{axodraw}

\makeatletter
\renewcommand{\fnum@table}{\textbf{\tablename~\thetable}}
\renewcommand{\fnum@figure}{\textbf{\figurename~\thefigure}}
\makeatother

\newlength{\myem}
\newcounter{mysubequation}[equation]

\makeatletter
\renewcommand{\section}{\@startsection{section}{1}{0em}%
        {-3.5ex \@plus -1ex \@minus -.2ex}%
        {2.3ex \@plus.2ex}%
        {\normalfont\large\bfseries}}
\renewcommand{\subsection}{\@startsection{subsection}{2}{0em}%
        {-3.25ex\@plus -1ex \@minus -.2ex}%
        {1.5ex \@plus .2ex}%
        {\normalfont\bfseries}}
\renewcommand{\subsubsection}%
        {\@startsection{subsubsection}{3}{0em}%
        {-3.25ex\@plus -1ex \@minus -.2ex}%
        {1.5ex \@plus .2ex}%
        {\normalfont\bfseries}}
\makeatother

\newcommand{\SNS}{Scuola Normale Superiore and INFN Sezione di Pisa, \\ 
Piazza dei Cavalieri 7, I--56126 Pisa, Italy}

\newcommand{\ICTP}{International Center for Theoretical Physics,
Strada Costiera 11, I-34000 Trieste, Italy}

\newcommand{\preprintnumber}{}
 
\newcommand{\titletext}{Consequences of Triplet Seesaw
for Leptogenesis} 
\newcommand{\authortext}{\large Thomas
  Hambye$^{\, a}$ and Goran Senjanovi\'c$^{\, b}$ 
\medskip\\\em\normalsize 
$\mbox{}^a$ \SNS
\\[0.1\baselineskip] 
$\mbox{}^b$ \ICTP}
\newcommand{\abstracttext}{We present the various leptogenesis scenarios
which may occur if, in addition to the ordinary 
heavy right-handed neutrinos, there exists a heavy scalar $SU(2)_L$
triplet coupled to leptons.
We show that the contributions of the right-handed neutrinos
and the 
triplet to the lepton asymmetry are proportional to their
respective
contributions to the neutrino mass matrix. 
A consequence of the triplet contribution to the lepton asymmetry is that 
there is no more upper bound on the neutrino masses 
from leptogenesis due to the fact that the neutrino mass constraints
do not necessarily induce asymmetry washout effects.  
We also show how such a triplet leptogenesis mechanism may emerge naturally
in the framework of the left-right symmetric theories, such as
Pati-Salam or $SO(10)$.}

\title{
\normalsize
\begin{tabular}[t]{l}\preprintnumber\end{tabular}
\vspace{1\baselineskip}\\
\Large\bfseries\titletext\bigskip}
\author{\begin{minipage}[t]{0.8\textwidth}
\normalsize\centering\authortext
\end{minipage}}
\date{}

\begin{document}

\bigskip
\maketitle
\begin{abstract}\normalsize\noindent
\abstracttext
\end{abstract}\normalsize\vspace{\baselineskip}


\section{Introduction}

Recent evidence for neutrino masses, and the belief that these masses 
are presumably associated to lepton number violation, have 
established the leptogenesis
mechanism \cite{fy} as today's favorite explanation of the 
baryon asymmetry of the universe. 
In the usual seesaw \cite{seesaw} mechanism 
the lepton asymmetry is generated
through the decay of heavy Majorana right-handed neutrinos, the same 
ones that lead to small neutrino masses. 
This scenario is particularly natural in the framework of 
theories that predict the existence of heavy right-handed neutrinos, 
such as $SO(10)$ \cite{georgi}, 
Pati-Salam \cite{ps} and
left-right symmetric theories \cite{ps,lr} in general. 

A possible loophole in the ordinary seesaw mechanism 
is that in the framework of these unified 
theories, the 
heavy neutrinos provide neither the only possible 
source of light neutrino masses
nor the only possible source of lepton asymmetry generation.
In a large class of renormalizable left-right symmetric theories, it 
is well known that the existence of triplet
Higgses also induces neutrino masses via the so-called type II 
seesaw mechanism \cite{mwms}.
This alternative can naturally provide a connection \cite{bsv} 
between
the 
large
atmospheric mixing and $b-\tau$ unification in the context of the
minimal supersymmetric $SO(10)$ theory \cite{soten}.

In this letter we analyze the effects of the triplet for the 
leptogenesis mechanism. We show that if the triplet type II seesaw dominates 
the neutrino masses, diagrams involving the triplet will also in
general dominate the 
leptogenesis. Even if the triplet is heavier than the lightest 
right-handed 
neutrino this will be the case via the decay of this lightest right-handed 
neutrino and a diagram involving a virtual triplet.
If instead, the triplet is lighter
than all the  right-handed neutrinos then it is the decay of the triplet
to two leptons involving virtual right-handed 
neutrinos which will
dominate.
The suggestion that the triplet could be important for leptogenesis, in
the context of the left-right model, was already made in Ref.~\cite{sod} 
where also the relevant diagrams were exhibited. The diagrams for the decay of
the right-handed neutrinos were also exhibited and estimated in 
Ref.~\cite{laz1}.
Here we 
calculate the corresponding CP asymmetries and relate them to the
neutrino mass constraints.

\section{The three types of lepton asymmetries}
\label{sec:triplet}

It is useful to start with the minimal situation where in addition to 
the SM particles there are three heavy right-handed neutrinos and one 
heavy $SU(2)_L$ Higgs triplet.
From the asymmetries we will obtain in this case, one can derive
easily the 
corresponding asymmetries in the more realistic left-right or $SO(10)$
models.
The relevant Lagrangian of this model reads:
\begin{eqnarray}
{\cal L} &\owns& -\frac{1}{2} M_{N_{i}} N^{T}_{Ri} C N_{Ri} 
-M^2_\Delta Tr\Delta_L^\dagger \Delta_L
- H^\dagger \bar{N}_{Ri}  (Y_N)_{ij} \psi_{jL} \nonumber \\
& & - (Y_\Delta)_{ij} \psi_{iL}^T C i \tau_2 
\Delta_L \psi_{jL} + \mu H^T i \tau_2 \Delta_L H + h.c. \,,
\label{lagr}
\end{eqnarray} 
with $\psi_{iL}= (\nu_{iL}$, $l_{iL})^T$, $H=(H^0,H^-)^T$ and 
\begin{equation}
\Delta_L=
\begin{pmatrix}
\frac{1}{\sqrt{2}}\delta^+ & \delta^{++}  \\
\delta^0 & - \frac{1}{\sqrt{2}} \delta^+ 
\end{pmatrix} \,.
\end{equation}
In the presence of these interactions, the triplet acquires a vev which will
be naturally seesaw suppressed if the triplet 
is heavy: $\langle \delta^0 \rangle = v_L \simeq \mu^\ast v^2/M^2_\Delta$.
Due to this vev there are now in general two sources of neutrino masses:
\begin{equation}
M_\nu= -Y_N^T M_N^{-1} Y_N v^2 + 2 Y_\Delta v_L
\label{Mnu}
\end{equation}
where the first term is the ordinary ``type I'' 
seesaw 
induced by the right-handed neutrinos and the second term is the 
triplet ``type II'' seesaw mass term, with $v=174$~GeV.

%
\begin{figure}[t]
\begin{center}
\begin{picture}(300,80)(0,0)
\Line(0,50)(30,50)
\DashArrowLine(60,80)(90,80){5}
\ArrowLine(60,20)(90,20)
\Line(60,80)(60,20)
\ArrowLine(60,80)(30,50)
\DashArrowLine(30,50)(60,20){5}
\Text(5,42)[]{$N_k$}
\Text(35,32)[]{$H$}
\Text(37,71)[]{$ l_l$}
\Text(69,50)[]{$N_j$}
\Text(85,72)[]{$H^\ast$}
\Text(85,28)[]{$ l_{i}$}
\Text(45,3)[]{(a)}
\Line(100,50)(120,50)
\DashArrowArc(135,50)(15,180,360){5}
\ArrowArc(135,50)(15,0,180)
\Line(150,50)(170,50)
\DashArrowLine(170,50)(200,80){5}
\ArrowLine(170,50)(200,20)
\Text(105,42)[]{$N_k$}
\Text(135,26)[]{$H$}
\Text(136,75)[]{$ l_l$}
\Text(160,42)[]{$N_j$}
\Text(187,75)[]{$H^\ast$}
\Text(187,24)[]{$ l_{i}$}
\Text(150,3)[]{(b)}
\Line(210,50)(240,50)
\DashArrowLine(270,80)(300,80){5}
\ArrowLine(270,20)(300,20)
\DashArrowLine(270,20)(270,80){5}
\DashArrowLine(240,50)(270,80){5}
\ArrowLine(270,20)(240,50)
\Text(215,42)[]{$N_k$}
\Text(245,32)[]{$l_l$}
\Text(247,71)[]{$H$}
\Text(282,50)[]{$\Delta_L$}
\Text(295,72)[]{$H^\ast$}
\Text(295,28)[]{$ l_{i}$}
\Text(255,3)[]{(c)}
\end{picture}
\end{center}
\caption{One-loop diagrams contributing to the asymmetry
from the $N_k$ decay.}
\label{fig}
\end{figure}
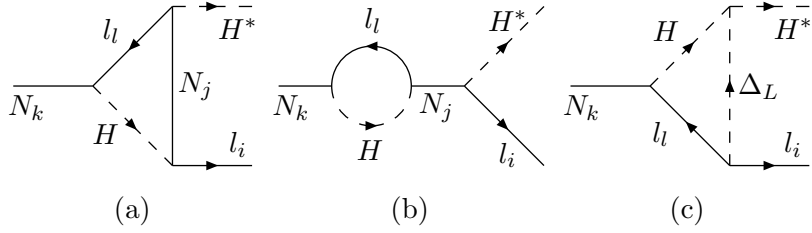

In this framework, depending on the values of the masses and couplings, 
the leptogenesis can be obtained either from the decay
of the right-handed neutrinos or from the decay of the triplet.
From the decay of the right-handed neutrinos to leptons and Higgs scalars
the CP asymmetry is given as usual by:
\begin{equation}
\varepsilon_{N_k}=\sum_i {{\Gamma (N_k \rightarrow l_i + H^*) - 
\Gamma (N_k \rightarrow 
\bar{l}_i + H)}\over{\Gamma (N_k \rightarrow l_i + H^*) +
\Gamma (N_k \rightarrow 
\bar{l}_i + H)}} \,.
\end{equation}   
This asymmetry is given by the interference of the ordinary tree level decay
with the 3 diagrams of Fig.~1. The first two diagrams are the ordinary 
vertex and self-energy diagrams involving another (virtual) right-handed 
neutrino and give
\begin{equation}
\varepsilon_{N_k}=\frac{1}{8 \pi} \sum_j
\frac{{\cal I}m[(Y_N Y_N^\dagger)_{kj}^2]}{\sum_i |(Y_N)_{ki}|^2}\sqrt{x_j}
\Big[1-(1+x_j) \log(1+1/x_j)+1/(1-x_j)\Big] \,,
\label{epsN}
\end{equation}
where $x_j=M^2_{N_j}/M^2_{N_k}$.
The third diagram of Fig.~1, which was already displayed without
calculations in Ref.~\cite{sod} and estimated in Ref.~\cite{laz1} involves 
a virtual triplet and is a new contribution.
Calculating it we obtain
\begin{equation}
\varepsilon_{N_k}^\Delta=-\frac{1}{2 \pi} \frac{\sum_j 
{\cal I}m[(Y_N)_{ki} 
(Y_N)_{kl} (Y^*_\Delta)_{il} \mu]}{\sum_i |(Y_N)_{ki}|^2 M_{N_k}} \, 
\Big(1-\frac{M^2_\Delta}{M^2_{N_k}} \log(1+M^2_{N_k}/M^2_\Delta) \Big) \,.
\label{epsND}
\end{equation}
Note that the tree level decay width is not affected by the existence of the
triplet:
\begin{equation}
\Gamma_{N_k}= \frac{1}{8 \pi} M_{N_k} \sum_i |(Y_N)_{ki}|^2 \,.
\label{gammaN}
\end{equation}

%
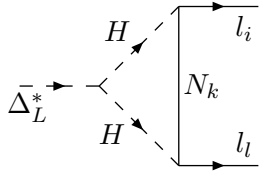
\begin{figure}[b]
\begin{center}
\begin{picture}(100,60)(0,0)
\DashArrowLine(0,30)(30,30){5}
\ArrowLine(60,60)(90,60)
\ArrowLine(60,0)(90,0)
\Line(60,60)(60,0)
\DashArrowLine(30,30)(60,60){5}
\DashArrowLine(30,30)(60,0){5}
\Text(3,22)[]{$\Delta_L^*$}
\Text(35,12)[]{$H$}
\Text(37,51)[]{$H$}
\Text(69,30)[]{$N_k$}
\Text(85,52)[]{$l_i$}
\Text(85,8)[]{$l_{l}$}
\end{picture}
\end{center}
\caption{One-loop diagram contributing to the asymmetry
from the $\Delta_L$ decay.}
\label{fig2}
\end{figure}

From the decay of the triplet to two leptons an asymmetry can also be produced.
It is given by the interference of the tree level process with the one-loop
vertex diagram, given in Fig.~2, involving a virtual right-handed neutrino 
\cite{sod}. 
Note that with one triplet alone there is no self-energy diagram, and 
therefore without at least one right-handed neutrino no asymmetry can be 
produced. At least two triplets are necessary in order to 
produce 
an asymmetry
without right-handed neutrinos, in which case the asymmetry comes from 
self-energy diagrams as was shown in Refs.~\cite{ms,hms} and also 
used in Ref.~\cite{laz2}. Here we will 
restrict ourselves to the case where there is only one $SU(2)_L$ 
triplet coupled to leptons (as it is 
in general the case in left-right and $SO(10)$ models, both ordinary
and supersymmetric).
Calculating the asymmetry from Fig.~2 we obtain:
\begin{eqnarray}
\varepsilon_\Delta &=& 2\cdot
{{\Gamma (\Delta_L^* \rightarrow l + l) - \Gamma (\Delta_L \rightarrow 
\bar{l} + \bar{l })}\over{\Gamma (\Delta_L^* \rightarrow l + l) 
    +\Gamma (\Delta_L \rightarrow \bar{l} + \bar{l}     )}}\\
&=& \frac{1}{8 \pi} \sum_k M_{N_k} \frac{\sum_{il}{\cal I}m[(Y_N^\ast)_{ki} 
(Y_N^\ast)_{kl}
(Y_\Delta)_{il} \mu^\ast]}{\sum_{ij}|(Y_\Delta)_{ij}|^2
 M^2_\Delta + |\mu|^2}
\log(1+M^2_\Delta/M^2_{N_k}) \,,
\label{epsD}
\end{eqnarray}
while the triplet decay width to two leptons and two scalar doublets
is given by:
\begin{equation}
\Gamma_\Delta=\frac{1}{8 \pi} M_\Delta \Big( \sum_{ij}|(Y_\Delta)_{ij}|^2 +
\frac{|\mu|^2}{M^2_\Delta} \Big) \,.
\label{gammaD}
\end{equation}
Note that there is such an asymmetry for each of the three components
of the triplet.
In the case where the lighter right-handed neutrino and the triplet
have approximately the same mass and same order of magnitude couplings,
all 3 types of asymmetries of Eqs.~(\ref{epsN}), (\ref{epsND}) 
and (\ref{epsD}) can play an 
important role. In the following we will
discuss the limiting cases where one process dominates over the others.
We will distinguish four such cases.

\subsection{Case 1: $M_{N_1} << M_\Delta$ with a dominant contribution of
the 
right-handed neutrinos to the light neutrino masses}

In the limit where the triplet couplings to two leptons are negligible 
with respect to the leading right-handed neutrino Yukawa couplings, and
with at least one right-handed neutrino much lighter than the triplet,
the triplet has a negligible effect for both the neutrino masses 
and the leptogenesis. This is equivalent to the ordinary right-handed
neutrino scenario without the triplet. Only the 2 diagrams of Fig.~1.a and
Fig.~1.b
have a non-negligible effect for leptogenesis. This case has been 
extensively studied in the 
literature (see e.g.~\cite{fy}, \cite{lpy}-\cite{afs})
and 
we have nothing to add here
to it.

\subsection{Case 2: $M_{N_1} << M_\Delta$ with a dominant triplet contribution 
to the light neutrino masses}

If $M_{N_1} << M_\Delta$, the decays of the right-handed
neutrino(s) will dominate 
the lepton asymmetry production.
Under the assumption that these neutrinos have a 
hierarchical pattern, only the decay of the lightest heavy neutrino $N_1$
is important for the asymmetry and Eq.~(\ref{epsN}) can be rewritten as 
\begin{equation}
\varepsilon_{N_1}=\frac{3}{16 \pi}\frac{M_{N_1}}{v^2}
\frac{\sum_{il}{\cal I}m[(Y_N)_{1i} (Y_N)_{1l}(M^{I*}_\nu)_{il}]}
{\sum_i |(Y_N)_{1i}|^2} \,,
\label{epsNbis}
\end{equation}
while under the assumption that the triplet is sizeably heavier than 
$N_1$, Eq.~(\ref{epsND}) can be rewritten as
\begin{equation}
\varepsilon_{N_1}^\Delta=-\frac{1}{8 \pi} \frac{M_{N_1}}{v^2} 
\frac{\sum_{il}{\cal I}m[(Y_N)_{1i} (Y_N)_{1l}(M^{II*}_\nu)_{il}]}
{\sum_i |(Y_N)_{1i}|^2} \,.
\label{epsNDbis}
\end{equation}
The above lepton 
asymmetries differ just by their respective contribution to the 
neutrino mass matrix (up to a factor $-2/3$).
As a result, if the triplet contribution dominates the neutrino 
mass matrix, it is very likely to dominate also the lepton asymmetry 
production through Eq.~(\ref{epsNDbis}).\footnote{There 
is however no one to one correspondence between the 
contribution to the neutrino masses and to leptogenesis
since the neutrino mass matrix is not a number but a 3 by 3 complex matrix. 
For example imagine
a case where both the $(Y_N)_{1j}$ as well as the $(Y_\Delta)_{ji}$
and $\mu$ 
have negligible phases with non-negligible phases for the $(Y_N)_{2j}$
and/or
$(Y_N)_{3j}$. Then the two diagrams of Fig.~1.a.b can still dominate 
the asymmetry production. However in this case one must be careful because,
if the $Y_N$ couplings have a
negligible contribution to the neutrino masses, the produced asymmetry
could be easily too small.}

An interesting feature of this scenario with respect to the one of 
case 1 is that since the 
couplings responsible
for the neutrino masses are not anymore responsible for the 
tree level decay, the neutrino mass constraints will not induce 
violation of the out-of-equilibrium condition 
\begin{equation}
\Gamma_{N_1} < H(T)|_{T=M_{N_1}}= \sqrt{\frac{4 \pi^3 g_\ast}{45}} 
\frac{T^2}{M_{Planck}}\Big|_{T=M_{N_1}}\,.
\label{ooecN}
\end{equation} 
where $g_\ast$ is the number of degrees of freedom active
at the temperature of the asymmetry production ($g_\ast \sim 100$).
Therefore the washout 
effects from inverse decays and 
associated scatterings will be avoided.\footnote{As well known (see
e.g.~\cite{buch}) in the type I 
seesaw model of neutrino masses and the corresponding 
leptogenesis, a value of $m_\nu$
of at least $\sqrt{\delta m^2_{atm}} \sim 0.05$~eV \cite{sk}
requires a
Yukawa coupling
which if associated to the decaying right-handed neutrino 
gives $\Gamma_{N_1}/H(M_{N_1}) \gtrsim 2 \cdot 10^{-3} M_{Planck} \sqrt{\delta
m^2_{atm}}/v^2 \sim 10-100$ 
independently of the mass of the right-handed neutrino,
resulting in suppression effects of similar order. These suppression
effects are not 
present in Eqs.~(\ref{epsND}) and (\ref{epsNDbis}).} 
As a consequence there is no upper bound on
the neutrino masses as in the usual type I seesaw
model.\footnote{This bound in
the usual scenario (i.e.~$m_\nu \lesssim 0.12$~eV \cite{buch})
is due to the fact that, the 
larger the (degenerate) light neutrino masses,
the 
larger have to be some of the $(Y_N)_{1j}$ couplings, which induce more
violation
of the 
out-of-equilibrium condition $\Gamma_{N_1} < H(M_{N_1})$ and hence more 
associated washout effects.
This obviously doesn't hold here.}
Moreover, the $Y_N$ couplings constrained 
by the out-of-equilibrium conditions for 
the decay width and scatterings can essentially cancel between the
numerator and the denominator in each of 
Eqs.~(\ref{epsND}) and (\ref{epsNDbis}),
leaving an asymmetry depending above all on the $Y_\Delta$ and $\mu$ couplings.
Since the triplet is heavier than $N_1$, the out-of-equilibrium
conditions on these couplings are much weaker and therefore there is 
essentially {\it no constraint} from these conditions on the size of the 
asymmetry.
Here, the larger the neutrino masses from the 
triplet, the larger the asymmetry and this is not accompanied
by larger washout effects as in the usual case.

It is also interesting to see in this case what happens to the upper
bound \cite{mur,th,di} on the asymmetry (or equivalently lower bound 
on the right-handed neutrino masses) due to the fact that neutrino 
masses are bounded from above.
From Eq.~(\ref{epsNDbis}) we see that such 
a bound still exists since the asymmetry 
is directly proportional to the neutrino mass matrix generated by the triplet.
If we assume a hierarchical pattern of 
light neutrino masses (i.e.~$m_\nu^{max} \simeq
\sqrt{\delta m^2_{atm}}\simeq 0.05$~eV \cite{sk}),
this bound is of the same order as in the case 1, 
i.e.~$M_{N_1} > \hbox{few} \, 10^{8}$~GeV.
Note that 
in this case where the triplet 
diagram dominates the asymmetry, this bound is an absolute bound unlike
for the pure right-handed neutrino case, because in the (single) 
triplet case there
is no self-energy diagram as the one of Fig.~1.b (which allows one 
to
avoid 
this bound through a resonance behavior if the right-handed 
neutrino masses are degenerate \cite{fps,pil}).
Therefore the triplet doesn't allow much progress
in being able to lower substantially 
the leptogenesis scale, which would be welcome for 
the eventual gravitino problem.\footnote{Note 
that, except if there are large cancellations between the type I and type II
seesaw contributions to the neutrino masses, this 
bound on $M_{N_1}$ still holds in the case of a mixed scenario
where both type I and type II contributions would be important for 
both neutrino masses and leptogenesis as long as the right-handed neutrinos
are not closely degenerate in mass.} The precise calculation of the 
asymmetries and of the corresponding $M_{N_1}$ lower limit in the 
supersymmetric case is 
called for and is left for a future publication. 

In order to show quantitatively 
that leptogenesis can be easily generated in this case and in order 
to illustrate
the above discussion, as 
an example, let us consider the case where the Yukawa couplings
$(Y_N)_{1j}$ of
$N_1$ gives $K_{N_1}=\Gamma_{N_1}/ H(M_{N_1}) \sim 1/10$. 
In this case the out-of-equilibrium condition is well satisfied and 
there is no sizeable washout effect associated to the $(Y_N)_{1j}$
couplings.
The contribution of the $(Y_N)_{1j}$ couplings 
to the neutrino masses
does not
exceed $M_\nu \sim (Y_N)_{1j}^2 v^2/M_{N_1}\sim 
(10^{-12}\,\hbox{GeV}) \, K_{N_1} \sim 10^{-4}$~eV
and is therefore negligible 
with respect to $\sqrt{\delta m^2_{atm}}\simeq 0.05$~eV \cite{sk} 
and $\sqrt{\delta m^2_{sol}}\simeq 0.008$~eV \cite{kam}. 
Let us assume in addition for simplicity
that the Yukawa couplings of the heavier 
right-handed neutrinos $N_{2,3}$ are also relatively small
and that the triplet is responsible for the neutrino 
masses (with for example $M_\Delta \sim (10-1000) \, M_{N_1}$).
In this case there is no substantial washout from diagrams involving the 
triplet. This is due to the fact that the triplet doesn't couple directly 
to $N_1$ and therefore 
there are no scatterings with a virtual triplet which enter 
into the Boltzmann equation ruling the number density of $N_1$.
The only potentially dangerous scatterings
enter into
the Boltzmann equation ruling the lepton number density, for 
instance scatterings such as 
$l + l \rightarrow \Delta \rightarrow H + H$.
These however are not relevant since, on the one hand, the
on-shell triplet contribution is already largely Boltzmann suppressed
at the temperature of the $N_1$ decay, and since on the other hand, the 
off-shell triplet contribution
is also suppressed by mass and couplings factors.
As a result, in this example, it is not necessary to consider
the explicit Boltzmann equations for the calculation of the produced
lepton 
asymmetry. The asymmetry will be given safely by $n_L/s \sim
\varepsilon_{N_1}^\Delta / g_\ast$.
For example the values 
$M_{N_1}=10^{10}$ GeV, 
$\mu=10^{11}$ GeV, 
$M_\Delta=  10^{12}$~GeV, 
$(Y_N)_{1j}^{max} = 2 \cdot 10^{-4}$, 
$(Y_\Delta)_{il}^{max} =10^{-2}$ 
give $m_\nu^{max} \simeq \sqrt{\delta m^2_{atm}}$ 
and $\varepsilon_N^\Delta \simeq
10^{-6}$ assuming a maximal phase. Taking $g_\ast \simeq 10^2$
we obtain an asymmetry as large as $n_L/s \simeq 10^{-8}$ which gives more
than enough freedom 
to get the needed value $n_L/s \sim 10^{-10}$ \cite{wmap,fisa}(by reducing
the phase for example).
In this numerical example, an increase in $m_{\nu}^{max}$ (resulting from
increasing the
$Y_\Delta$ and $\mu$ couplings) results simply in a linear increase 
of the produced lepton asymmetry.

Finally note that all the discussion above for this case remains true
even for $M_\Delta \sim M_{N_1}$, as long as the triplet contribution
dominates the neutrino masses as we assumed here. In particular,
except if there are cancellations in the interplay of phases,
$\varepsilon_{N_1}^\Delta$ remains dominant over $\varepsilon_{N_1}$ 
and even
$\varepsilon_\Delta$. The only additional restriction is that 
the relation $\Gamma_\Delta <
H(M_\Delta)$ will be violated by at least a factor $\sim 10-100$
due to neutrino mass constraints 
(i.e.~$\Gamma_\Delta/H(M_\Delta) \gtrsim
2 \cdot 10^{-3} M_{Planck} \sqrt{\delta m^2_{atm}}/v^2$). 
This
induces a washout from scatterings such as 
$l+l \rightarrow \Delta \rightarrow H+H$ which will be effective 
at the temperature of the $M_{N_1}$ decay,
but this
washout will be quite moderate because  
it intervenes only in the Boltzmann equation of the 
lepton number density, not in the one of the $N_1$ number density.
This leaves enough freedom to have successful leptogenesis
even for neutrino masses well above the other cosmological bounds
\cite{wmap}.

\subsection{Case 3:  $M_{N_1} >> M_\Delta$ with a dominant right-handed 
neutrino contribution to the neutrino masses}

Let us now consider a possibility that the triplet is much lighter 
than all the right-handed 
neutrinos. 
In this case leptogenesis will be dominated
by the decay of the triplet to two leptons and the one loop diagram 
is the one of Fig.~2. 
If in addition the type I seesaw mechanism dominates the neutrino masses,
the asymmetry of Eq.~(\ref{epsD}) becomes
\begin{equation}
\varepsilon_\Delta =-\frac{1}{8 \pi}
\frac{M^2_\Delta}{ (\sum_{ij} |(Y_\Delta)_{ij}|^2 M^2_\Delta +|\mu|^2)} 
\frac{1}{v^2}
{\cal I}m[(M^{I\ast}_\nu)_{il} (Y_\Delta)_{il} \mu^*]
\,.
\label{epsDbis}
\end{equation}
This can perfectly well lead to the needed lepton asymmetry (this 
case being obtained from case 2 by inverting the role of
the right-handed neutrinos and the triplet).
As in the case 2, there is no upper bound on the neutrino masses 
from leptogenesis because 
the couplings responsible for the neutrino masses are not responsible
for the tree level decay. Moreover, also as in 
the case 2, the couplings constrained 
by the out-of-equilibrium conditions for 
the decay width and associated scatterings can essentially cancel between 
the numerator and denominator in each of Eqs.~(\ref{epsD}) and (\ref{epsDbis}),
leaving an asymmetry depending above all on the $Y_N$ couplings which 
are much less constrained by the out-of-equilibrium conditions.
Note however that, in contrast to the case 2, the decaying 
particle is not a $SU(2)_L \times
U(1)$
gauge singlet. Therefore there is the additional constraint that 
the triplet has not to be kept in thermal equilibrium by the gauge 
scatterings down to a temperature sizeably below its mass (otherwise an 
asymmetry
can be created only from a Boltzmann suppressed number density of triplets).
This puts the lower limit on $M_\Delta$ above the lower limit on
$M_N$ in the previous cases.

To illustrate this, one can
consider a parameter
configuration just opposite to the one of the example in the case 2,
i.e.~with suppressed triplet couplings and larger 
right-handed neutrino Yukawa couplings. Considering such a configuration, with
for example  $M_\Delta = 10^{13}$~GeV, one can show
that a large enough 
lepton asymmetry can be produced. 
We estimate that a successful leptogenesis can also occur 
for $M_\Delta$ down to $\sim 10^{11-12}$~ GeV \cite{hms,olive} 
for light neutrino masses
below a few tenth of eV.
An explicit calculation,  
beyond the scope of 
the present letter,
with full Boltzmann equations would be necessary
to 
see exactly down to which value 
of the triplet and right-handed neutrino masses 
leptogenesis can be successfully generated. 

Note also that a lower bound on the triplet mass still holds here 
from neutrino constraints, in a similar way as in the case 2,
but this one is not useful as it is below the lower bound on the triplet mass
from the gauge scattering washout effects.

Note finally that if the right-handed neutrinos dominate the
neutrino masses as in this case, even for $M_{N_1} \sim M_\Delta$,
$\varepsilon_\Delta$ will still be important for leptogenesis as
long as the triplet mass is large enough to avoid the gauge
scattering effects. In this case $\varepsilon_\Delta$
is of order $\varepsilon_{N_1}$ and both dominate over 
$\varepsilon_N^\Delta$, see Eqs.~(\ref{epsN}),
(\ref{epsND}) and (\ref{epsD}). The
interplay of
Boltzmann equations is quite involved in this case.

\subsection{Case 4: $M_{N_1} >> M_\Delta$ case with dominance of the
triplet 
contribution to the neutrino masses}

In this case too an asymmetry can be produced
from the decay of the triplet 
to two leptons, i.e. the diagram of Fig.~2 and Eq.~(\ref{epsD}). 
In terms of the neutrino mass matrix generated by the triplet,
with $M_\Delta << M_{N_1}$, $\varepsilon_\Delta$ can be rewritten as
\begin{equation}
\varepsilon_\Delta =\frac{1}{16 \pi}
\frac{M^2_\Delta}{M_{N_k} v^2}
\frac{
{\cal I}m[(M^{II}_\nu)_{il} (Y_N^\ast)_{ki} (Y_N^\ast)_{kl}]
}{ (\sum_{ij} |(Y_\Delta)_{ij}|^2
+\frac{|\mu|^2}{M^2_\Delta} )} 
\,.
\label{epscase4}
\end{equation}
There are quite a few constraints here:
\begin{itemize}
\item First, in this case the asymmetry will be proportional to the $Y_N$
couplings wich do not dominate the neutrino masses, and in this sense it is
suppressed with respect to all other cases; compare Eq.~(\ref{epscase4})
with Eqs.~(\ref{epsNDbis}) and (\ref{epsDbis}).
\item Second, as in the case 3, the triplet
needs to have a mass large enough to avoid large damping effects 
from gauge scatterings.
\item In addition, 
due to the fact that the couplings 
responsible for the decay width
are also responsible for the neutrino masses, 
the condition $\Gamma_\Delta < H(M_\Delta)$ will be
violated by at least a factor $\sim 10-100$ due to the atmospheric
neutrino constraints, similarly to the case 1 discussed in the footnote 2
(i.e.~$\Gamma_\Delta/H(M_\Delta) \gtrsim
2 \cdot 10^{-3} M_{Planck} \sqrt{\delta m^2_{atm}}/v^2$). 
In fact for this condition there is even less
freedom than in the case 1 because in the case 1 the condition
$\Gamma_{N_1} <
H(M_{N_1})$ can still be satisfied 
in the hierarchical case or inverse hierarchical case by assuming that
$N_1$ has small
Yukawa 
couplings (the heaviest light 
neutrino masses being induced by $N_{2,3}$).
Since there is only one
triplet, we don't have this freedom and this condition is always
violated.
\item Moreover, 
due to this fact that the couplings 
responsible for the decay width
are also responsible for the neutrino masses, 
there is now as in the 
case 1 an upper bound on neutrino masses. 
\end{itemize}
Altogether this leads to a case which can still successfully
generate 
the leptogenesis although within a narrower range of parameters
than the other cases. The upper bound on neutrino masses and the lower
bound on $M_\Delta$  depend highly on how negligible we assume
the type I seesaw contribution for the neutrino masses. They result from
a rather involved interplay of Boltzmann equations beyond 
the scope of this letter.

\section{Implications from and for the left-right model}

A heavy $SU(2)_L$ triplet $\Delta_L$ emerges naturally in any 
left-right symmetric theory with a renormalizable see-saw mechanism.
We discuss this generic feature in a prototype $SU(2)_L \times SU(2)_R
\times U(1)_{B-L}$ model, but it equally well applies to theories such as
Pati-Salam or $SO(10)$. In this theory, the leptons are left and
right-handed doublets $\psi_{iL,R}=(\nu_i, l_i)^T_{L,R}$ and the analog of the
standard model Higgs $H$ is a bi-doublet $\Phi$ (consisting of two
$SU(2)_L$ doublets). The right-handed neutrinos receive their masses
through the $SU(2)_R$ triplet $\Delta_R$; this in turn, due to $L-R$
symmetry necessarily implies the existence of our $SU(2)_L$ triplet
$\Delta_L$. The relevant Yukawa and Higgs potential terms which reproduce
the Lagrangian of Eq.~(\ref{lagr}) now read as
\begin{eqnarray}
{\cal L} \owns &-&(Y_\Delta)_{ij} \big[ \psi_{iL}^T C i \tau_2 \Delta_L
\psi_{jL}+ \psi_{iR}^T C i \tau_2 \Delta_R \psi_{jR} \big] \nonumber \\
&-& Y_{ij}^{(1)} \bar{\psi}_{iR} \Phi_1 \psi_{jL} 
- Y_{ij}^{(2)} \bar{\psi}_{iR} \Phi_2 \psi_{jL} \nonumber \\
&-& \lambda_{ij} Tr \Delta_R^\dagger \Phi_i \Delta_L \Phi_j^\dagger \, +
\, h.c. 
\label{lagrLR}
\end{eqnarray}
with
\begin{equation}
  \label{eq:PD}
 \Phi= 
\begin{pmatrix}
\phi_1^0 & \phi_1^+  \\
-\phi_2^- & \phi_2^0 
\end{pmatrix},
\,\,\,\,\,
\Delta_{L,R}=
{\begin{pmatrix}
\frac{1}{\sqrt{2}}\delta^+ & \delta^{++}  \\
\delta^0 & - \frac{1}{\sqrt{2}} \delta^+ 
\end{pmatrix}}_{L,R}\,,
\end{equation}
and $\Phi_1= \Phi$, $\Phi_2= \tau_2 \Phi^\ast \tau_2$.
In writing Eq.~(\ref{lagrLR}) we have used the following
$SU(2)_L \times SU(2)_R$ transformations 
properties: $\psi_{L,R} \rightarrow U_{L,R} \psi_{L,R}$; $\Delta_{L,R}
\rightarrow U_{L,R} \Delta_{L,R} U^\dagger_{L,R}$; 
$\Phi \rightarrow U_R \Phi U^\dagger_L$. We have also used the left-right
symmetry, i.e.~symmetry under: $\psi_L \leftrightarrow \psi_R$, 
$\Delta_L \leftrightarrow \Delta_R$, $\Phi \leftrightarrow \Phi^\dagger$.

Denote, next, $\langle \phi_i^0 \rangle = v_i$, and introduce the notation
$H_i=(\phi_i^0, \phi_i^-)^T$. It is easy to show that $H=(v_1 H_1+ v_2
H_2)/\sqrt{v_1^2+v_2^2}$ corresponds to the standard model Higgs doublet,
whereas $H'=(v_2 H_1- v_1 H_2)/\sqrt{v_1^2+v_2^2}$ has zero vev 
and picks up a mass proportional to $M_R=\langle \Delta_R \rangle$
and decouples from the low energy sector.
Comparison
with 
Eq.~(\ref{lagr}), says that after diagonalizing $Y_\Delta$,
\begin{equation}
M_{N_i}= 2 (Y_{\Delta}^{diag})_{ii} M_R
\end{equation}
gives Dirac neutrino Yukawa couplings as a function of $Y^{(1)}$ and
$Y^{(2)}$
\begin{equation}
Y_N={{Y^{(1)} v_1 + Y^{(2)} v_2} \over {\sqrt{v_1^2 + v_2^2}}} \,,
\end{equation}
and
\begin{equation}
\mu=\frac{
(\lambda_{11}+\lambda_{22}) v_1 v_2 + \lambda_{12} v_2^2 + \lambda_{21}
v_1^2
}{
v_1^2+v_2^2
}\, M_R \,.
\end{equation}
The main message is very simple. For large $M_R$, we have effectively the
situation as in section 2, but with one important proviso for the type II
seesaw: right and left-handed neutrino masses are proportional to each
other (given by $Y_\Delta$ and the $\langle \Delta_R \rangle$ and
$\langle \Delta_L \rangle$
respectively) and so are their eigenvalues.
This must be taken into account for a quantitative analysis of the
asymmetry.

Three more comments are in order:
\begin{itemize}
\item First, in the $SU(2)_L \times SU(2)_R
\times U(1)_{B-L}$ theory the right-handed neutrino masses, and
$M_{{\Delta}_L}$ 
and $\mu \sim \lambda \langle \Delta_R \rangle$ are clearly not predictable
and there is no way of 
knowing which
particle is lightest and therefore responsible for leptogenesis.
A natural prejudice, encouraged by what we know in the standard model, 
is a hierarchy of $M_{N_i}$ and not so light scalar 
particles, (i.e.~$M_{\Delta_L} >> M_{N_1}$). This, to us, makes
the cases 1 and 2 more
plausible than 3 and 4, although not necessarily correct.
Keep also in mind that a heavy $\Delta_L$ can still naturally dominate 
the light neutrino masses; after all Yukawa couplings $Y_N$ could be
much smaller than $Y_\Delta$. In this sense the case 2 is not less
plausible
than the case 1.
\item Second, we don't know the mass of the second doublet $H'$ in the
bi-doublet $\Phi$. As we said, unless we fine tune it, it is naturally of
order $M_R$, but it is not clear whether it lies above or below the mass of
decaying particle responsible for the leptogenesis.
We assume here that $H'$ is heavy enough not to affect
leptogenesis. We do this only for the reason of simplicity,
but it
is straightforward to generalize it to the case of a lighter  $H'$; one
must only take into account the fact that $\mu$ becomes a $2 \times 2$
matrix and keep an index $i$ on $Y_N^{(i)}$, $i=1,2$.
\item Third, for the case where $M_{N_1} < M_\Delta$, we assume 
that the $SU(2)_R$ gauge bosons are heavy enough to
avoid too large damping effects from the corresponding gauge scatterings.
To this end, it has been estimated that these interactions 
are out-of-equilibrium for
$M_{W_R} > (2 \cdot 10^{5}\, \hbox{GeV})(M_{N_1}/10^2 \, 
\hbox{GeV})^{3/4}$ \cite{usss}. This should 
be the object of a more detailed study. It is 
likely that for specific sets of parameters, leptogenesis can be successfully
generated for even smaller values of $M_{W_R}$.
\end{itemize}

Let us see now  what happens to $\varepsilon_{N_1}^\Delta$ in this
theory. From $(Y_\Delta)_{ij}=\frac{1}{2} \delta_{ij}
\frac{M_{N_i}}{M_R}$ in the basis of diagonal $M_N$, one obtains from
Eq.~(\ref{epsND}) for the case of
hierarchical $N_i$ and for $M_\Delta >> M_{N_1}$ (i.e.~in the, to us,
most interesting case 2):
\begin{equation}
\varepsilon_{N_1}^\Delta \simeq - \frac{1}{8 \pi} 
\sum_j \frac{M_{N_1}M_{N_j}}{M^2_\Delta M_R} 
\frac{
{\cal I}m [ (Y_N)_{1j} (Y_N)_{1j} \mu ]
}{\sum_i |(Y_N)_{1i}|^2} \,. 
\end{equation}
This can be contrasted with $\varepsilon_{N_1}$ in the same limit
\begin{equation}
 \varepsilon_{N_1} \simeq - \frac{3}{16 \pi} \sum_j 
\frac{M_{N_1}}{M_{N_j}} 
\frac{
{\cal I}m [ \sum_j (Y_N Y_N^\dagger)_{1j}^2 ] 
}{\sum_i |(Y_N)_{1i}|^2} \,.
\end{equation}
Let us take for illustration the same example of the section 2
with hierarchical masses for the $N_i$, say 
$M_{N_1}\simeq 10^{10}$~GeV,
$M_{N_2}\simeq 10^{12}$~GeV,
$M_{N_3}\simeq 10^{13}$~GeV and $M_R\simeq 10^{15}$~GeV.
Clearly for $Y_N^2 v^2/M_N \lesssim \sqrt{\delta m^2_{sol}} \sim 0.008$~eV 
as dictated by the dominance of the triplet
seesaw, one gets $\varepsilon_{N_1} \lesssim 10^{-7}$ to be compared 
with $\varepsilon_{N_1}^\Delta \simeq 10^{-6}$.

Last but not least, note that the lower 
bound $M_{N_1} \gtrsim \hbox{few}\, 10^8$ GeV
in the case of hierarchical right-handed neutrinos translates in a lower limit
about one order of magnitude higher on the scale of $SU(2)_R$ symmetry 
breaking $M_R$, or in 
other words on the mass of the right-handed gauge bosons.

\section{Summary}

The analysis of leptogenesis and neutrino masses with a triplet in
addition to right-handed neutrinos is somewhat involved but still, a
clear picture emerges. 
If $M_\Delta > M_{N_1}$, 
the triplet will have in general 
a non-negligible 
contribution to the leptogenesis as soon as it has a non-negligible 
contribution to the neutrino masses, even if the triplet is much heavier than
$N_1$ (see the case 2).
If $M_\Delta < M_{N_1}$, with $M_\Delta \gtrsim 10^{11-12}$~GeV,
even with
rather small 
couplings (i.e.~not necessarily dominating the neutrino masses), the triplet 
diagrams will in general dominate the production
of the asymmetry (see cases 3 and 4).
If $M_\Delta < M_{N_1}$, with $M_\Delta \lesssim 10^{11-12}$~GeV, the 
triplet will not be able to produce the asymmetry due to 
gauge scattering effects and it must have in addition tiny couplings
not to wash out the asymmetry (which in this case can be produced
by the pure right-handed neutrino contribution).
The situation which definitely allows most freedom in the parameter
space 
is the one of the case 2 with triplet seesaw neutrino masses and
leptogenesis
from
the decay of the right-handed neutrino(s) (i.e.~with a triplet mass
larger or of order the lightest right-handed neutrino mass). 
The most constrained situation is with the dominance of the triplet for the
neutrino masses and a triplet mass lighter than all right-handed neutrino
masses (case 4).

One of the consequence of the triplet contribution to leptogenesis is that
there is no more an upper limit on neutrino masses from leptogenesis (see
the cases 2 and 3). This is due to the fact that in $\varepsilon_N^\Delta$
and $\varepsilon_\Delta$, the couplings responsible for the tree
level decay are not necessarily the ones dominating the neutrino masses.
Therefore the neutrino mass
constraints do not necessarily imply violation of the out-of-equilibrium
conditions on the decay width and associated scatterings, even 
for degenerate light neutrino masses (see the cases 2 and 3).
Moreover these couplings responsible for the tree level decay can
essentially cancel in the numerator and denominator of the asymmetry,
leaving an asymmetry depending above all on other couplings
much less constrained by the out-of-equilibrium conditions.
The size of the asymmetry is bounded from above only by the size
of the neutrino masses (which results in the lower bound on $M_{N_1}$
similar to the one of the usual case without the 
triplet, i.e.~$M_{N_1} \gtrsim \hbox{few} \, 10^{8}$~GeV).

In left-right symmetric theories with the seesaw mechanism, the existence
of the triplet $\Delta_L$ is a must if one sticks to renormalizability.
Regarding neutrino masses, it is a priori impossible to know wether they
originate from the triplet or right-handed neutrinos (or both).
In this sense it is equally probable that the triplet or right-handed
neutrinos play an important role for neutrino masses and therefore for
leptogenesis. It is still to be seen what happens in more constrained
theories such as the minimal renormalizable $SO(10)$ theory, where the
small number of parameters allows one in principle to find out the
origin
of the seesaw mechanism and in turn the origin of the lepton asymmetry.

\vspace{1.1cm}
\begin{center}{\large Acknowledgments}
\end{center}  
It is a pleasure to thank Utpal Sarkar for useful discussions several
years ago. We are also
grateful to Pasquale Di Bari, Pavel Fileviez Perez,
Alejandra Melfo and Francesco Vissani for discussions and comments. We 
appreciate the warm hospitality of the CMS group
at the Technical University of Split (FESB), where part of this work was done. 
This work was 
supported by the TMR, EC-contract No.~HPRN-CT-2000-00148
and HPRN-CT-2000-00152.

\end{document}